\documentclass[12pt,english,floatfix,superscriptaddress,aps,prd,preprint,showkeys]{revtex4}
\usepackage[utf8]{inputenc}
\usepackage{amsmath}
\usepackage{amssymb}
\usepackage{amsbsy}
\usepackage{amsfonts}
\usepackage{amsopn}
\usepackage{amstext}
\usepackage{graphicx}
\usepackage{amssymb}
\usepackage{amsfonts}
\usepackage{amsmath}
\usepackage{amsmath,amsthm,amsfonts,amssymb}
\usepackage[mathcal]{eucal}
\usepackage{mathrsfs}
\usepackage{graphicx}
\usepackage[english]{babel}
\usepackage{color}
\usepackage{esint}
\usepackage[dvips]{epsfig}
\usepackage[dvips]{graphicx}
\usepackage{float}
\usepackage{units}
\usepackage{textcomp}

\usepackage{hyperref}
\usepackage{slashed}

\newcommand{\ie}{\begin{equation}}
\newcommand{\fe}{\end{equation}}
\newcommand{\se}{\begin{eqnarray}}
\newcommand{\ff}{\end{eqnarray}}

\begin{document}

\title{Topological and noninertial effects in an Aharonov-Bohm ring}


\author{R. R. S. Oliveira}
\email{rubensfq1900@gmail.com}
\affiliation{Universidade Federal do Cear\'a (UFC), Departamento de F\'isica,\\ Campus do Pici, Fortaleza - CE, C.P. 6030, 60455-760 - Brazil.}

\date{\today}

\begin{abstract}

In this paper, we study the influence of topological and noninertial effects on a Dirac particle confined in an Aharonov-Bohm (AB) ring. Next, we explicitly determine the Dirac spinor and the energy spectrum for the relativistic bound states. We observe that this spectrum depends on the quantum number $n$, magnetic flux $\Phi$ of the ring, angular velocity $\omega$ associated to the noninertial effects of a rotating frame, and on the deficit angle $\eta$ associated to the topological effects of a cosmic string. We verified that this spectrum is a periodic function and grows in values as a function of $n$, $\Phi$, $\omega$, and $\eta$. In the nonrelativistic limit, we obtain the equation of motion for the particle, where now the topological effects are generated by a conic space. However, unlike relativistic case, the spectrum of this equation depends linearly on the velocity $\omega$ and decreases in values as a function of $\omega$. Comparing our results with other works, we note that our problem generalizes some particular cases of the literature. For instance, in the absence of the topological and noninertial effects ($\eta=1$ and $\omega=0$) we recover the usual spectrum of a particle confined in an AB ring ($\Phi\neq{0}$) or in an 1D quantum ring ($\Phi=0$).

\end{abstract}


\maketitle


\section{Introduction}

In recent years, the study of particles confined in quantum rings (QRs) gained much attention due to its possible technological applications, such as in sources and photonic detectors \cite{Michler}, single-photon emitters, single-electron transistors, nano-flash memories \cite{Nowozin}, qubits for spintronics and quantum computing, magnetic RAM memories and storage media \cite{Fomin}, etc. In particular, QRs also constitute a unique scenario for the study of some quantum-mechanical phenomena \cite{Ihn2003}, for instance, in Aharonov-Bohm oscillations \cite{Webb,Aichinger}, persistent currents \cite{Ihn,Viefers,Splettstoesser}, Berry phases \cite{Lopes}, and Rashba spin–orbit interaction \cite{Bulgakov}. Moreover, QRs are divided into two classes, namely, the one-dimensional rings or 1D rings (rings of constant radius) \cite{Buttiker,Lorke,Fuhrer,Meijer,Bolivar} and the two-dimensional rings or 2D rings (rings of variable radius)\cite{Tan,Bulaev,Duque,Nowak,B}.

According to the literature, a specific case of 1D QRs which has been gaining relevance in recent years are the so-called Aharonov-Bohm (AB) rings \cite{Lorke,Fuhrer,Citro,Nitta}. In this way, to model AB rings is necessary to combine the dynamics of 1D QRs with the AB effect. Currently, AB rings were studied in different physical contexts, such as in the Aharonov-Casher effect \cite{Citro,Nitta}, Lorentz symmetry violation \cite{Belich}, mesoscopic decoherence \cite{Hansen}, electromagnetic resonator \cite{Reulet}, Rashba spin-orbit interaction \cite{Aeberhard,Shelykh}, quantum asymmetry \cite{Pedersen}, persistent currents \cite{Yi}, quantum transport \cite{Frustaglia}, etc. Posteriorly, it was formulated a relativistic AB ring model for Dirac particles; however, the nonrelativistic limit (low-energy regime) of this ring not was analyzed and much less compared with the usual results of literature \cite{Cotaescu}. Recently, AB rings were studied in de Sitter expanding universe \cite{Cotaescu2}, and their thermodynamic properties were calculated theoretically in the high and low-energy regime \cite{Oliveira}.

On the other hand, the study of noninertial effects due to rotating frames also have been widely investigated since to 1910 decade \cite{Matsuo2011}, where the best-known effects are the Sagnac \cite{Sagnac,Post}, Barnett \cite{Barnett,Ono} and Mashhoon \cite{Mashhoon} effects. In condensed matter physics, noninertial effects were investigated in the quantum Hall effect \cite{Fischer,Matsuo}, Bose-Einstein condensates \cite{Schweikhard,Cooper,Bretin}, fullerene molecules \cite{Shen,Lima}, and in atomic gases \cite{Cooper2008,Lu}. Now, in the high-energy regime, noninertial effects also gained focus of investigations in some quantum systems described by relativistic wave equations \cite{Wang2018,Santos,Hosseinpour2015,Castro}. For instance, considering the Dirac equation (DE) in a rotating frame, this equation is applied in problems involving spin currents \cite{Dayi2018,MatsuoPRL}, Dirac oscillator and cosmic strings \cite{Bakke2012}, fullerene molecules \cite{Cavalcante,Gonzalez,Kolesnikov}, nanotubes and carbon nanocons \cite{Cunha,Gomes}, Sagnac and Hall effects \cite{Anandan,Zubkov}, and external magnetic fields \cite{Liu,Chernodub2017}.

The present paper has as its goal to investigate the influence of topological and noninertial effects on the relativistic and nonrelativistic quantum dynamics of a Dirac particle confined in an AB ring. In particular, to achieve this goal, we use mainly Refs. \cite{Bakke2012,Greiner}. To include the topological and noninertial effects in our problem, we apply a rotation in the coordinate system in form $\varphi\to\varphi+\omega t$, where $\omega$ is the constant angular velocity of the rotating frame, and we use the cosmic string background (linear gravitational topological defect) described by a deficit angle $\eta$, where $\eta\equiv 1-4G\mu/c^2$ and $\mu$ is the linear mass density of the cosmic string \cite{Bakke2012,Vilenkin}. It is important to mention that the first papers that studied the dynamics of a nonrelativistic 2D QR in a rotating frame with and without the presence of a topological defect are found in Refs. \cite{Dantas,Roncaglia}. Already in Ref. \cite{Fonseca}, is analyzed the noninertial effects on an atom with magnetic quadrupole moment confined in a 2D QR. Last but not least, the topological and noninertial effects analogous to the of this paper were recently applied in another type of quantum nanostructure, the so-called quantum dots \cite{Rojas}.

This work is organized as follows. In Sect. \ref{sec2}, we introduce the cosmic string spacetime and the configuration of the minimal electromagnetic coupling in a rotating frame. In Sect. \ref{sec3}, we investigate the influence of topological and noninertial effects on the relativistic quantum dynamics of a charged Dirac particle confined in an AB ring. Next, we explicitly determine the Dirac spinor and the energy spectrum for the relativistic bound states of the particle. In Sect. \ref{sec4}, we analyze the nonrelativistic limit of our results. In addition, we compare also our problem with other works, where we observe that our results generalizes some particular cases of the literature. In Sect. \ref{conclusion}, we present our conclusions. In this work, we use the natural units where $\hbar=c=G=1$ and the spacetime with signature $(+ - - -)$.

\section{The rotating magnetic cosmic string background \label{sec2}}

In this section, we describe the curved spacetime background in a rotating frame. The chosen curved spacetime is the cosmic string spacetime along of the $Z$-axis, where the line element in cylindrical coordinates for this object is given by \cite{Bakke2012,Vilenkin}
\ie ds^2=dT^2-dR^2-\eta^2R^2 d\Phi^2-dZ^2,
\label{lineelement}\fe
where the deficit angle $\eta$ is defined in the range $0<\eta<1$ and the geometry characterized by this line element has a conical singularity that gives rise to a curvature centered on the cosmic string axis ($Z$-axis), however, in all other places the curvature is null \cite{Bakke2012}. This conical singularity is represented by the following curvature tensor
\ie R^{\rho,\varphi}_{\rho,\varphi}=\frac{1-\eta}{4\eta}\delta_2(\bf r),
\label{curvature}\fe
where $\delta_2(\bf r)$ is the 2D Dirac delta. In condensed matter physics, it is already well known that linear topological defects as dislocations and disclinations can be described by a line element in the same way that a topological defect in general relativity \cite{Katanaev,Kleinert}. It is worth mentioning that in the case of the cosmic string, the spatial part of its line element corresponds to the line element of a disclination. Therefore, when we take the nonrelativistic limit of the DE, we can extend this formalism to the solid state physics context \cite{Bakke2012}.

As we are interested in working with a rotating frame, we should use the following coordinate transformations
\ie T=t, \ \ R=\rho, \ \ \Phi=\varphi+\omega t, \ \ Z=z,
\label{coordinate}\fe
where $\omega$ is the constant angular velocity (not an angular frequency) of the rotating frame. So, with the transformations (\ref{coordinate}), the line element (\ref{lineelement}) becomes \cite{Bakke2012}
\ie ds^2=(1-\kappa^2)dt^2-2\kappa\eta\rho d\varphi dt-d\rho^2-\eta^2\rho^2d\varphi^2-dz^2,
\label{lineelement2}\fe
where o parameter $\kappa$ is defined as $\kappa\equiv\omega\eta\rho$.

With the line element (\ref{lineelement2}), we need to construct a local reference frame where the observers will be placed; consequently, we can to define from this the Dirac matrices in the rotating curved spacetime background. In this way, a local reference frame can be built through of an noncoordinate basis given by $\hat{\theta}^a=e^a_{\ \mu}(x)dx^\mu$, which its components $e^a_{\ \mu}(x)$ satisfy the following relation \cite{Bakke2012}
\ie \mathrm{g}_{\mu\nu}(x)=e^a_{\ \mu}(x)e^b_{\ \nu}(x)\eta_{ab},
\label{tensor}\fe
where $g_{\mu\nu}(x)$ is the curved metric tensor, $\eta_{ab}$ is the Minkowski metric tensor and the indices $a,b=0,1,2,3$ indicate the local reference frame. The components of the noncoordinate basis $e^a_{\ \mu}(x)$ are called tetrads or vierbein, whose inverse is written as $dx^\mu=e_{\ a}^\mu(x)\hat{\theta}^a$, where $e^{a}_{\ \mu}(x)e^{\mu}_{\ b}(x)=\delta^{a}_{\ \ b}$ and $e^{\mu}_{\ a}(x)e^{a}_{\ \nu}(x)=\delta^{\mu}_{\ \nu}(x)$ must be satisfied. Our interest is to build a rotating frame where there is no torque on the system. Thus, we choose the tetrads and and its inverse in the form
\ie e^{a}_{\ \mu}(x)=\left(\begin{array}{cccc}
\sqrt{1-\kappa^2} & 0 & -\frac{\kappa\eta\rho}{\sqrt{1-\kappa^2}} & 0	\\
0 & 1 & 0 & 0	\\
0 & 0 & \frac{\eta\rho}{\sqrt{1-\kappa^2}} & 0	\\
0 & 0 & 0 & 1
\end{array}\right), \ \ e^{\mu}_{\ a}(x)=\left(\begin{array}{cccc}
\frac{1}{\sqrt{1-\kappa^2}} & 0 & \frac{\kappa}{\sqrt{1-\kappa^2}} & 0	\\
0 & 1 & 0 & 0 \\
0 & 0 & \frac{\sqrt{1-\kappa^2}}{\eta\rho} & 0	\\
0 & 0 & 0 & 1
\end{array}\right).
\label{tetrads} \fe

It should be noted that these choices make the 1, 2 and 3-axis of the local reference frame be parallel to the $\rho$, $\varphi$ and $z$-axis of the curved spacetime, respectively. With the informations about the choice of the local reference frame, we can obtain now the one-form connection $\omega^a_{\ b}=\omega^{\ a}_{\mu\ b}(x)dx^\mu$ through the Maurer-Cartan structure equations. In the absence of torsion (or torque), these equations are written as \cite{Bakke2012}
\ie d\hat{\theta}^a+\omega^a_{\ b}\wedge\hat{\theta}^b=0,
\label{Maurer-Cartan}\fe
where the operator $d$ is the exterior derivative and the symbol $\wedge$ means the external product. Therefore, the non-null components of the one-form connection are \cite{Bakke2012}
\ie \omega^{\ 0}_{t\ 1}(x)=\omega^{\ 1}_{t\ 0}(x)=-\frac{\kappa\omega\eta}{\sqrt{1-\kappa^2}},
\label{one-form}\fe
\ie \omega^{\ 1}_{t\ 2}(x)=-\omega^{\ 2}_{t\ 1}(x)=-\frac{\omega\eta}{\sqrt{1-\kappa^2}},
\label{two-form}\fe
\ie \omega^{\ 0}_{\rho\ 2}(x)=\omega^2_{\rho\ 0}(x)=\frac{\omega\eta}{(1-\kappa^2)},
\label{there-form}\fe
\ie \omega^{\ 0}_{\varphi\ 1}(x)=\omega^{\ 1}_{\varphi\ 0}(x)=-\frac{\kappa\eta}{\sqrt{1-\kappa^2}},
\label{four-form}\fe
\ie \omega^{\ 1}_{\varphi\ 2}(x)=-\omega^{\ 2}_{\varphi\ 1}(x)=-\frac{\eta}{\sqrt{1-\kappa^2}}.
\label{five-form}\fe

From now, we focus our attention on the configuration of the vector potential of the AB effect in the cosmic string background in a rotating frame. To insert a magnetic interaction into DE due to a spin-1/2 particle with electric charge $q$ ($q<0$) we should introduce into DE a minimal coupling given by $i\gamma^{\mu}(x)\partial_\mu\to i\gamma^{\mu}(x)(\partial_\mu+iqA_\mu(x))$, where $A_\mu(x)=e^a_{\ \mu}(x)A_a$ is the electromagnetic four-vector and $\gamma^{\mu}(x)$ are the curved gamma matrices \cite{Bakke2012,Greiner}. Explicitly, $A_\mu(x)$ is written in the rotating frame of the observer $(\omega\neq0)$ as follow
\ie A_\mu(x)=\left(0,0,-\frac{\eta\rho}{\sqrt{1-\kappa^2}}A_\varphi,0\right), \ \ (A_0=A_\rho=A_z=0),
\label{four-vector}\fe
where $A_\varphi$ is the angular component of the electromagnetic field $A_a=(0,0,-A_\varphi,0)$ written in the rest inertial frame of the observer $(\omega=0)$. In particular, the AB effect is generated by an infinite solenoid of radius $a$ perpendicular to the polar plane and whose vector potential is given by $\vec{A}=\frac{\Phi}{2\pi\eta\rho}\hat{e}_\varphi$, where $\Phi>0$ is the constant magnetic flux in the interior of the solenoid \cite{rubens,Aharonov,Griffiths}. On the other hand, the parameter $\eta$ arises because the solenoid coincides with the axis of symmetry of the cosmic string \cite{Bakke2012}.

\section{Relativistic quantum dynamics of an Aharonov-Bohm ring in the rotating magnetic cosmic string background \label{sec3}}

In this section, we determine the relativistic bound-state solutions of a charged Dirac particle confined in an AB ring under the influence of topological and noninertial effects. We start our discussion initially from a system in cylindrical coordinates, and posteriorly, we turn the system in a dynamics purely planar with constant radius (1D system). In that way, the covariant $(3+1)$-dimensional DE in the curved spacetime for a charged particle with rest mass $m_0$ in the presence of an external electromagnetic field reads as follows \cite{Bakke2012,Greiner}
\ie i\gamma^{\mu}(x)[\nabla_\mu(x)+iqA_{\mu}(x)]\Psi(t,{\bf r})=m_0\Psi(t,{\bf r}),
\label{1}\fe
where $\gamma^{\mu}(x)=e^\mu_{\ a}(x)\gamma^a$ are the curved gamma matrices and $\gamma^a$ are the gamma matrices defined in the inertial Minkowski spacetime, $\nabla_\mu(x)=\partial_\mu+\Gamma_\mu(x)$ is the covariant derived, being $\Gamma_\mu(x)=\frac{i}{4}\omega_{\mu ab}(x)\sigma^{ab}$ the spinorial connection and the quantity $\omega_{\mu ab}(x)$ is the spin connection. Besides that, with the expressions
\eqref{one-form}, \eqref{two-form}, \eqref{there-form}, \eqref{four-form} and \eqref{five-form}, we obtain
\ie \Gamma_t(x)=-\frac{1}{2}\frac{\kappa\omega\eta}{\sqrt{1-\kappa^2}}\alpha^1-\frac{i}{2}\frac{\omega\eta}{\sqrt{1-\kappa^2}}\Sigma^3
\label{spinorial1},\fe
\ie \Gamma_\rho(x)=\frac{1}{2}\frac{\omega\eta}{(1-\kappa^2)}\alpha^2
\label{spinorial2},\fe
\ie \Gamma_\varphi(x)=-\frac{1}{2}\frac{\kappa\eta}{\sqrt{1-\kappa^2}}\alpha^1-\frac{i}{2}\frac{\eta}{\sqrt{1-\kappa^2}}\Sigma^3
\label{spinorial3},\fe
where implies that
\ie \gamma^\mu(x)\Gamma_\mu(x)=\frac{1}{2}\frac{i\eta\gamma^0\omega}{(1-\kappa^2)^{3/2}}\Sigma^3+\frac{1}{2\rho}\gamma^1
\label{spinorial}.\fe

With respect to curved gamma matrices, we have
\ie \gamma^0(x)=\frac{1}{\sqrt{1-\kappa^2}}\gamma^0+\frac{\kappa}{\sqrt{1-\kappa^2}}\gamma^2
\label{gama0},\fe
\ie \gamma^1(x)=\frac{1}{\sqrt{1-\kappa^2}}\gamma^1
\label{gama1},\fe
\ie \gamma^2(x)=\frac{\sqrt{1-\kappa^2}}{\eta\rho}\gamma^2
\label{gama2},\fe
\ie \gamma^3(x)=\gamma^3
\label{gama3}.\fe

In the polar coordinates system where $\partial_z\Psi(t,{\bf r})=0$, we can rewrite Eq. \eqref{1} as follows
\ie A\Psi(t,\rho,\varphi)=m_0\Psi(t,\rho,\varphi),
\label{2}\fe
where the operator $A$ is defined in the form
\ie A\equiv\frac{i}{\sqrt{1-\kappa^2}}\left[(\gamma^0+\kappa\gamma^2)\partial_t+\gamma^1\partial_\rho\right]+\gamma^2\left[i\frac{\sqrt{1-\kappa^2}}{\eta\rho}\partial_\varphi+q A_\varphi\right]+\frac{i}{2\rho}\gamma^1-\frac{\eta\gamma^0}{(1-\kappa^2)^{3/2}}\vec{S}\cdot\vec{\omega},
\label{3}\fe
being the term $\vec{S}\cdot\vec{\omega}$ so-called spin-rotation coupling \cite{Mashhoon,Hehl}, where $\vec{\omega}=\omega\hat{e}_z$ and $\vec{S}=\frac{1}{2}\vec{\Sigma}$ is the spin operator \cite{Greiner}.    

We see that it is difficult to proceed without the simplification of Eq. (\ref{2}). So, to exactly solve Eq. (\ref{2}), we consider the linear velocity of the rotating frame being small compared with the velocity of light; thus, we have $\kappa^2\ll 1$ \cite{Bakke2012}. Therefore, using this condition, the configuration of $A_\varphi$, and the following conditions to model an 1D QR: $\partial_\rho\to-\frac{1}{2\rho}$ and $\rho\to R$, where $R=const.$ is the radius of the ring \cite{Costa}, Eq. \eqref{2} becomes
\ie i\frac{\partial\Psi(t,\varphi)}{\partial t}=H_{ring}\Psi(t,\varphi),
\label{4}\fe
where
\ie H_{ring}=-\alpha^2\left[\frac{1}{\eta R}\left(i\frac{\partial}{\partial\varphi}-\frac{\Phi}{\Phi_0}\right)+i\omega\eta R\frac{\partial}{\partial t}\right]+\eta\vec{S}\cdot\vec{\omega}+\gamma^0 m_0.
\label{5}\fe
being $H_{ring}$ the Dirac Hamiltonian for the rotating AB ring, $\Phi_0\equiv\frac{2\pi}{e}$ is the elementary magnetic flux quantum and we assume that $q=-e$ ($e>0$).

Since we are working now in a $(1+1)$-dimensional spacetime, it is convenient to write the Dirac matrices $\vec{\alpha}=(0,\alpha^2,0)=(0,-\alpha_2,0)$ and $\gamma^0$ and the spin operator $\vec{S}=(0,0,\frac{1}{2}\Sigma^3)$ in terms of the $2\times 2$ Pauli matrices, i.e., $\alpha_2=\sigma_2$ and $\gamma^0=\Sigma^3=\sigma_3$ \cite{Oliveira,Greiner,rubens}. In this way, setting the following ansatz for the two-component Dirac spinor \cite{Oliveira,Greiner}
\ie \Psi(t,\varphi)=e^{-iEt}\left(
           \begin{array}{c}
            \psi(\varphi) \\
            \chi(\varphi) \\
           \end{array}
         \right)
\label{spinor},\fe
we obtain from \eqref{4} a system of two first-order coupled differential equations given by
\ie\left(E-m_0-\frac{\eta\omega}{2}\right)\psi(\varphi)=\left[\frac{1}{\eta R}\left(\frac{d}{d\varphi}+i\frac{\Phi}{\Phi_0}\right)-im_0\Omega R\right]\chi(\varphi),
\label{6}\fe
\ie\left(E+m_0+\frac{\eta\omega}{2}\right)\chi(\varphi)=\left[-\frac{1}{\eta R}\left(\frac{d}{d\varphi}+i\frac{\Phi}{\Phi_0}\right)+im_0\Omega R\right]\psi(\varphi),
\label{7}\fe
where $\Omega\equiv\frac{\omega\eta E}{m_0}$ is an effective angular frequency, and $E$ is the total relativistic energy.

Substituting \eqref{7} into \eqref{6}, we get the following linear differential equation with constant coefficients
\ie \left[\frac{d^{2}}{d\varphi^{2}}+2iB\frac{d}{d\varphi}+\mathcal{E}\right]\psi(\varphi)=0,
\label{8}\fe
where
\ie B\equiv\left(\frac{\Phi}{\Phi_0}-m_0 \eta\Omega R^2\right), \ \ \mathcal{E}\equiv (\eta R)^2\left[E^2-\left(m_0+\frac{\eta\omega}{2}\right)^2\right]-B^2.
\label{coefficient}\fe

It is easy to verify that the solution of \eqref{8} is given in the form \cite{Oliveira,Griffiths}
\ie\psi(\varphi)=C e^{i\lambda\varphi}, 
\label{solution}\fe
where $C>0$ is a normalization constant. Consequently, the coefficient $\lambda$ takes the form
\ie \lambda_\pm=-B\pm(\eta R)\sqrt{E^2-\left(m_0+\frac{\eta\omega}{2}\right)^2}.
\label{coefficient2}\fe

However, as $\psi(\varphi)$ satisfies the following periodicity condition $\psi(\varphi+2\pi)=\psi(\varphi)$, implies that $\lambda_\pm$ must be equal to the a integer $n$ ($n=0,\pm 1,\pm 2,\ldots$). Therefore, using this condition, we have the following relativistic energy spectrum of a Dirac particle confined in an AB ring under the influence of topological and noninertial effects
\ie E^\sigma_{n}=-\omega\left(n+\frac{\Phi}{\Phi_0}\right)+\sigma\sqrt{\left[\omega\left(n+\frac{\Phi}{\Phi_0}\right)\right]^2+\left(m_0+\frac{\eta\omega}{2}\right)^2+\left[\frac{1}{\eta R}\left(n+\frac{\Phi}{\Phi_0}\right)\right]^2},
\label{spectrum}\fe
where $n$ is a quantum number, being that $n>0$ describe a particle moving in the same sense that the electric current of the solenoid and $n<0$ describe a particle moving in the opposite sense \cite{Griffiths}, $\sigma=+1$ corresponds to the positive energy states (particle), $\sigma=-1$ corresponds to the negative energy states (antiparticle), and we also use the condition $(\omega\eta R)^2\ll 1$ (how should be) for construct the spectrum. We see that the spectrum (\ref{spectrum}) explicitly depends on the magnetic flux $\Phi$ that models the AB ring, angular velocity $\omega$ of the rotating frame, and on the deficit angle $\eta$ associated to the topological effects of the cosmic string. In particular, we have that $E^\sigma_{n}(\Phi\pm\Phi_0)=E^\sigma_{n\pm 1}(\Phi)$, i.e., the energy spectrum is a periodic function with periodicity $\pm\Phi_0$ \cite{V}. On the other hand, we note that for $\sigma=-1$ (with $n>0$) the terms of the spectrum \eqref{spectrum} are summed, while for $\sigma=+1$, the terms are subtracted, consequently, the energies of the antiparticle (in absolute values) are larger that of the particle. Moreover, we note also that the quantum number $n$ and the parameters $\omega$, $\Phi$ and $\eta$ have the function of increasing the energies of the spectrum, i.e., in the limits $\eta\to 0$ (extremely dense cosmic string) and $n=\omega=\Phi\to\infty$, we have $\vert E_{n}^\sigma\vert\to\infty$. Now, comparing the spectrum (\ref{spectrum}) with the literature, we verified that in the absence of the topological and noninertial effects ($\eta=1$ and $\omega=0$) we recover the usual spectrum of a Dirac particle confined in a relativistic AB ring \cite{Cotaescu,Oliveira}.

From here on let us concentrate on the original form of the two-component Dirac spinor for the relativistic bound states of the system. Therefore, substituting the function \eqref{solution} into Eq. \eqref{7}, and immediately such result in \eqref{spinor}, we obtain the following Dirac spinor
\begin{equation}
\Psi_n(t,\varphi)=C_n e^{i(n\varphi-Et)}\left(
           \begin{array}{c}
            1 \\
             -\frac{i}{\nu}\left(n+\frac{\Phi}{\Phi_0}+m_0\eta\Omega R^2\right) \\
           \end{array}
         \right),
\label{spinor2} 
\end{equation}
where $\nu\equiv \eta R\left(E+m_0+\frac{\eta\omega}{2}\right)$.

According to Ref. \cite{Greiner}, the constant $C_n$ is determined by following normalization condition 
\ie \int_0^{2\pi}\Psi^\dagger_n(t,\varphi)\Psi_n(t,\varphi) d\varphi=1.
\label{I}\fe

Thus, the constant $C_n$ it's written explicitly as
\ie C_n=\frac{1}{\sqrt{2\pi\left[1+\frac{1}{\nu^2}\left(n+\frac{\Phi}{\Phi_0}+m_0\eta\Omega R^2\right)^2\right]}}.
\label{C2}\fe

\section{Nonrelativistic limit\label{sec4}}

In this section, we analyze the nonrelativistic limit of our results. To get this limit is necessary to consider that most of the total energy of the system stay concentrated in the rest energy of the particle \cite{Greiner}, i.e., $E\cong\varepsilon+m_0$, where $m_0\gg\varepsilon$ and $m_0\gg\omega$. In this way, using this prescription and the condition $(\eta\omega R)^2\ll 1$ in Eq. \eqref{8}, we get the following Schr\"odinger equation (SE) for a spinless particle confined in a nonrelativistic AB ring under the influence of topological and noninertial effects
\ie \left[\frac{d^{2}}{d\varphi^{2}}+2i\beta\frac{d}{d\varphi}+\epsilon\right]\psi(\varphi)=0,
\label{limit}\fe
where
\ie \beta\equiv\left(\frac{\Phi}{\Phi_0}-m_0\omega(\eta R)^2\right), \ \ \epsilon\equiv(\eta R)^2\left[\varepsilon-\frac{\eta\omega}{2}\right]-\beta^2.
\label{limit2}\fe

However, in this low-energy regime, the topological effects are now generated by a conic Euclidean space \cite{Andrade}. In addition, we note that in the absence of the topological and noninertial effects ($\eta=1$ and $\omega=0$), we recover the usual SE for a particle confined in an AB ring ($\Phi\neq 0$) \cite{Aeberhard,Griffiths} or in an 1D QR ($\Phi=0$) \cite{Viefers}.

Now, using the prescription $E\cong\varepsilon+m_0$ with $(\eta\omega R)^2\ll 1$ in \eqref{spectrum}, we have the following nonrelativistic energy spectrum of a particle confined in an AB ring under the influence of topological and noninertial effects
\ie \varepsilon_n=-\omega\left(n+\frac{\Phi}{\Phi_0}-\frac{\eta}{2}\right)+\frac{1}{2m_0}\left[\frac{1}{\eta R}\left(n+\frac{\Phi}{\Phi_0}\right)\right]^2.
\label{spectrum2}\fe

We see that the nonrelativistic spectrum \eqref{spectrum2} also is a periodic function with periodicity $\pm\Phi_0$ and and grows in values as a function on the quantum number $n$, magnetic flux $\Phi$, and of the deficit angle $\eta$ associated to the topological effects of the conic space, i.e., in the limits $\eta\to 0$ and $n=\Phi\to\infty$, we have $\varepsilon_n\to\infty$. However, unlike relativistic case, the spectrum \eqref{spectrum2} depends linearly on the velocity $\omega$ and decreases in values as a function of $\omega$. Now, comparing the spectrum \eqref{spectrum2} with the literature, we verified that in the absence of the noninertial effects ($\omega=0$), we recover the usual spectrum of a particle confined in an AB ring with and without the influence of topological effects \cite{Viefers,Fuhrer,Belich,Aeberhard,Ihn,Griffiths}. Last but not least, in the absence of the topological and noninertial effects ($\eta\to 1$ and $\omega=0$) with  $\Phi=0$, we recover the usual spectrum of a particle confined in an 1D QR \cite{Viefers}.

\section{Conclusion\label{conclusion}}

In this paper, we study the influence of topological and noninertial effects on the relativistic and nonrelativistic quantum dynamics of a Dirac particle confined in an AB ring. To include the topological and noninertial effects in our problem, we apply a constant rotation in the system, and we use the cosmic string background. Besides that, to model the AB ring, we work in the polar coordinate system whose radial coordinate is a constant. Next, we determine the relativistic bound-state solutions of the system, given by two-component Dirac spinor and the energy spectrum. In particular, we observe that this spectrum depends on the quantum number $n$, magnetic flux $\Phi$ of the ring, angular velocity $\omega$ associated to the noninertial effects of a rotating frame, and on the deficit angle $\eta$ associated to the topological effects of a cosmic string. We also note that this spectrum is a periodic function with periodicity $\pm\Phi_0$, where $\Phi_0$ is the elementary magnetic flux quantum, and grows in values as a function of $n$, $\Phi$, $\omega$ and $\eta$. On the other hand, we verified that the energies of the antiparticle (in absolute values) are larger that of the particle. Now, comparing our relativistic spectrum with other works, we note that our problem generalizes a particular case of the literature, i.e., in the absence of the topological and noninertial effects ($\eta=1$ and $\omega=0$) we recover the usual spectrum of a Dirac particle confined in an relativistic AB ring.

Finally, we study the nonrelativistic limit of our results. For instance, considering that the most of the total energy of the system stay concentrated in the rest energy of the particle, we obtain the SE for a spinless particle confined in an AB ring under the influence of topological and noninertial effects, where now these topological effects are generated by a conic space. Besides, we note that in the absence of the topological and noninertial effects ($\eta=1$ and $\omega=0$), we recover the usual SE for a particle confined in an AB ring ($\Phi\neq{0}$) or in an 1D QR ($\Phi=0$). With respect to the nonrelativistic energy spectrum, we note that this spectrum also is a periodic function with periodicity $\pm\Phi_0$ and grows in values as a function of $n$, $\Phi$ and $\eta$. However, unlike relativistic case, this spectrum depends linearly on the velocity $\omega$ and decreases in values with a function of $\omega$. Now, comparing also our nonrelativistic spectrum with other works, we note that our problem generalizes some particular cases of the literature, i.e., in the absence of the noninertial effects ($\omega=0$) we recover the usual spectrum of a particle confined in an AB ring with and without the influence of topological effects, while for $\eta=1$ and $\omega=\Phi=0$, we recover the spectrum of a particle confined in an 1D QR.

\section*{Acknowledgments}
\hspace{0.5cm} The author would like to thank the Conselho Nacional de Desenvolvimento Cient\'{\i}fico e Tecnol\'{o}gico (CNPq) for financial support.

\end{document}